\documentclass[prd, aps, showpacs]{revtex4}

\usepackage{latexsym, graphicx}

\begin{document}

\title{Backreaction and Unruh effect:\\
New insights from exact solutions of uniformly accelerated detectors}

\author{Shih-Yuin Lin}
\email{sylin@phys.cts.nthu.edu.tw}
\affiliation{
Physics Division, National Center for Theoretical Sciences,
P.O. Box 2-131, Hsinchu 30013, Taiwan}
\author{B. L. Hu}
\email{blhu@umd.edu} \affiliation{Maryland Center for Fundamental
Physics, Department of Physics, University of Maryland, College Park,
Maryland 20742-4111, USA}
\date{25 July 2007}

\begin{abstract}
Using nonperturbative results obtained recently for an uniformly
accelerated Unruh-DeWitt detector, we discover new features      
in the dynamical evolution of the detector's internal degree of
freedom, and identified the Unruh effect derived originally from
time-dependent perturbation theory as operative  
in the ultra-weak coupling and ultra-high acceleration limits. The
mutual interaction between the detector and the field engenders
entanglement between them, and tracing out the field leads to a mixed
state of the detector even for a detector at rest in Minkowski
vacuum.
Our findings based on this exact solution shows clearly the
differences from the ordinary result where the quantum field's
backreaction is ignored in that the detector no longer behaves like a
perfect thermometer.
From a calculation of the evolution of the reduced density matrix of
the detector, we find  that the transition probability from the
initial ground state over an infinitely long duration of interaction
derived from time-dependent perturbation theory is existent in the
exact solution only in transient under special limiting conditions
corresponding to the Markovian regime. Furthermore, the detector at
late times never sees an exact Boltzmann distribution over the energy
eigenstates of the free detector, thus in the non-Markovian regime
covering a wider range of parameters the Unruh temperature cannot be
identified inside the detector.
\end{abstract}


\maketitle

\section{Introduction}

The Unruh effect states that an observer while undergoing uniform
acceleration in the Minkowski vacuum feels as if it lives in a thermal
state at the Unruh temperature. It is usually demonstrated by way of
time-dependent perturbation theory (TDPT) for a ``particle detector"
over an infinitely long duration of interaction \cite{Unr76, DeW79, BD}.
Consider the detector as a quantum mechanical object with internal
degree of freedom $Q$ coupling to a quantum field $\Phi$ by the
interaction Hamiltonian
\begin{equation}
  H_I = \lambda_0 Q(\tau) \Phi(z^\mu(\tau))
\end{equation}
where $\lambda_0$ is the coupling constant, $\tau$ is the proper
time for the detector and $z^\mu(\tau)$ is the trajectory of the
uniformly accelerated detector (UAD) with proper acceleration $a$.
Suppose initially the detector-field system is in a product state,
i.e., they are uncorrelated,
\begin{equation}
  |\left.\tau_0 \to -\infty \right> = |\left. E_0\right> \otimes
  |\left. 0_M\right> , \label{initstat}
\end{equation}
where $|\left. E_0\right>$ is the ground state of the free
detector and $|\left. 0_M\right>$ is the Minkowski vacuum of the
free field. From TDPT,  the transition probability $P_{0\to n}$
from the ground state to the $n$-th excited state $|\left.
E_n\right>$ of the detector is given to first order in
$\gamma\equiv\lambda_0^2/8\pi m_0$, by \cite{BD}
\begin{eqnarray}
  P_{0\to n} &=& {1\over \hbar^2}\sum_{n_{\rm k}}\left|
    \left< E_n,n_k\right.| \int_{-\infty}^\infty d\tau H_I(\tau)
    e^{i(E_n-E_0)\tau/\hbar}|\left. E_0, 0_M \right>\right|^2 \nonumber\\
  &=& {\lambda_0^2\over \hbar^2}\left|\left< E_n\right.| Q(0)
    |\left. E_0\right>\right|^2 {\cal R}\left({E_n-E_0\over\hbar}\right),
   \label{Planck}
\end{eqnarray}
which is nonzero. Here
\begin{equation}
  {\cal R}(\kappa) = \int_{-\infty}^\infty d\tau d\tau'
    e^{-i\kappa(\tau-\tau')} \left<\right. 0_M | \Phi(z(\tau))
    \Phi(z(\tau')) |0_M\left.\right> = {\hbar \over 2\pi}
  {\kappa \eta \over e^{2\pi \kappa/a}-1}. \label{respfn}
\end{equation}
is the response function with $\eta \equiv \int_{-\infty}^\infty
d\tau$ being the duration of interaction \footnote{By definition,
Eq.(58) in \cite{LH2005} with $\kappa'=\kappa$ is actually the
response function ${\cal R}(-\kappa)/ (2\pi)^2$ with $\eta \equiv
2\pi\delta(0)$.}. In particular, for a harmonic oscillator with
renormalized natural frequency $\Omega_r$, all the transition
probabilities with $n>1$ are $O(\gamma^2)$, so the only
non-vanishing $P$ of $O(\gamma)$ is
\begin{equation}
  P_{0\to 1} = {\lambda_0^2\over 4\pi  m_0} {\eta \over
   e^{2\pi \Omega_r /a} -1}
\label{Psho}
\end{equation}
From the Planck factor in $(\ref{Psho})$ one can identify the
Unruh temperature $T_U = \hbar a/2\pi k_B$. When $a\to 0$, the
transition probability per unit time $P_{0\to 1}/\eta$ vanishes,
which implies that there is no excitation in an inertial detector
initially prepared in its ground state \cite{DeW79}.

According to the above calculation of the transition probability
one may expect that after the coupling is switched on, the reduced
density matrix of the detector would evolve continuously from the
ground state to higher excited states, with the detector finally
ending up in equilibrium with a Boltzmann distribution $\sim \exp
[-E_n/(k_BT_U)]$ at the Unruh temperature $T_U$. Then one can read
off the Unruh temperature from the late-time distribution inside
the detector.

In this note we use the exact solutions to an Unruh-Dewitt detector
\cite{Unr76, DeW79} in uniform acceleration interacting with a
quantum field to show that the above described conventional scenario
is good only in the Markovian regime corresponding to the limits of
ultra-high acceleration or ultra-weak coupling. The transition
probability calculated from the infinite-time TDPT is valid only in
transient under restricted conditions. The evolution of the detector
under general conditions is quite different from the above picture.

The difference between this new understanding and the conventional
picture originates from the fact that the full interplay between the
detector and the field is included here whereas the conventional
approach based on  perturbation theory operative for infinite time
ignores the backreactions -- the use of TDPT over indefinite time
amounts to invoking a Markovian approximation which imposes rather
severe limitations. The conventional wisdom is built upon conditions
which cannot reflect the most general features in the full dynamics
of the detector-field system made possible here by the
nonperturbative solutions.

This paper is organized as follows: In Sec.II we consider a moving
Unruh-DeWitt detector in 4-D Minkowski space coupled to a quantum
scalar field $\Phi$ and calculate the reduced density matrix after
tracing out the field. In Sec.III we calculate the transition
probability of the detector from the ground state to an excited
state. We elaborate on the physical meaning of the two constants
$\Lambda_0, \Lambda_1$ in the theory. In Sec.IV we calculate the
purity and the von Neumann entropy or the entropy of entanglement of
the detector-field system and identify an effective temperature
$T_{\rm eff}$ which is ostensibly different from the Unruh
temperature. We identify the range of validity of the conventional
results to that which corresponds to making a Markovian approximation
with its strong limitations. In Sec.V we summarize our main results
with remarks pertaining to the issues raised in this paper and
conclude with a comment on how our nonperturbative results may bear
on a new understanding of the Hawking effect of black hole radiance.

\section{Reduced Density Matrix for Detector}

Consider an Unruh-DeWitt (UD) detector moving in (3+1) dimensional
Minkowski space. The total action is given by\cite{LH2005}
\begin{equation}
  S=\int d\tau {m_0\over 2}\left[ \left(\partial_\tau Q\right)^2
    -\Omega_0^2 Q^2\right] -\int d^4 x {1\over 2}\partial_\mu\Phi
    \partial^\mu\Phi  +{\lambda_0} \int d\tau \int d^4 x Q(\tau) \Phi (x)
  \delta^4\left(x^{\mu}-z^{\mu}(\tau)\right), \label{Stot1}
\end{equation}
where $Q$ is the internal degree of freedom of the detector,
assumed to be a harmonic oscillator with bare mass $m_0$ and bare natural
frequency $\Omega_0$. The scalar field $\Phi$ is assumed to be
massless, and $\lambda_0$ is the coupling constant.

This UD detector theory behaves like the quantum Brownian motion (QBM)
of a harmonic oscillator interacting with an Ohmic bath provided by
the 4-D scalar quantum field \cite{HM}.
The QBM model is a useful comparison because it shows clearly the
dissipative and stochastic behavior of the dynamics arising from
the interplay between the system and the environment, and the
influence functional treatment incorporates the backreaction of
the environment on the system (which could be either the quantum
field or the harmonic oscillator depending on what one is after)
in a self-consistent way. Since the QBM model indicates that there
are nontrivial activities at zero temperature \cite{HM, HPZ, UZ89},
we caution that even for the $a=0$ case the detector is not just
laying idle but has interesting physical features due to its
interaction with the vacuum fluctuations in the quantum field.

Unruh and Zurek \cite{UZ89} have studied an exactly solvable QBM
model where a harmonic oscillator interacts with a massless scalar
field in 2-D. They derived the exact master equation for the reduced
density matrix of the system (oscillator) at a temperature 
determined by the initial state of the field, and observed some
general features different from the conventional Markovian results
(valid for an ohmic bath at ultra-high temperature). 
One feature is the dependence of the UV cut-off in the master 
equation and the reduced density matrix,
and thus also in the von Neumann entropy of the system. When the interaction
is switched on, the factors in the master equation and the entropy
have initial ``jolts" over a very short time scale corresponding to 
the UV cut-off frequency, which cause significant change in the
coherence of the quantum state of the system, while the coherence
residing in each subsystem is rapidly transferred into the correlation
between them. This effect is discussed in detail in \cite{HPZ}.
Below we will see a similar behavior occurring in UD detector theory
beyond the Markovian regime. 

Suppose the initial state of the system at $\tau_0$ is given by
$(\ref{initstat})$, a direct product of the ground state for $Q$
and the Minkowski vacuum for $\Phi$. In the Schr\"odinger
representation, this initial state is a product of Gaussian
functions,
\begin{equation}
  \psi_0 (\tau_0) = N \exp \left[ - A Q^2 - \int {d^3k\over (2\pi)^3}
  {d^3k'\over (2\pi)^3} B^{\rm k k'}  \Phi_{\rm k} \Phi_{\rm k'}\right].
\end{equation}
Owing to the linearity of the model,
this state must evolve into a general Gaussian form
\begin{equation}
  \psi_0 (\tau) = N(\tau) \exp \left[ - A(\tau) Q^2 -
   B^{\rm k k'}(\tau)\Phi_{\rm k} \Phi_{\rm k'}
  -2C^{\rm k}(\tau)\Phi_{\rm k} Q \right] ,
\end{equation}
after the coupling is switched on. Here
we have used the DeWitt notation where each upper-lower indices
pairing indicates an integration (or summation). So the density
matrix for this pure state in the $(Q,\Phi_{\rm k})$
representation can be written as $\rho [Q,\Phi_{\rm
k},Q',\Phi'_{\rm k};\tau]= \psi_0[Q,\Phi_{\rm k};\tau]\psi_0^*
[Q',\Phi'_{\rm k};\tau]$, and the reduced density matrix defined
by tracing out the field $\Phi_{\rm k}$ reads
\begin{equation}
  \rho^R (Q,Q';\tau) 
  =\int {\cal D}\Phi_{\rm k}
  \psi_0 [Q,\Phi_{\rm k};\tau] \psi_0^* [Q',\Phi_{\rm k};\tau] =
  \exp \left[ -G^{ij}(\tau)Q_i Q_j - F(\tau)\right],\label{RDM0}
\end{equation}
where $i,j =1,2$, $Q_i = (Q, Q')$. The factors $G^{ij}$ and $F$
could be obtained by solving the master equations \cite{FHR}. But
here we do not need to solve them directly, because the two-point
functions of the detector have been obtained in the Heisenberg
picture in Ref.\cite{LH2005}. (These results are placed in
Appendix A for convenience.)
One can reconstruct the complete evolution of the reduced density
matrix using those two-point functions of the detector by solving the
simple algebraic relations
\begin{eqnarray}
  G^{11}+G^{22}+2G^{12} &=& {1\over 2 \left<\right. Q^2
    \left.\right>}, \label{G1}\\
  G^{11}+G^{22}-2G^{12} &=& {2\over \hbar^2\left<\right. Q^2
    \left.\right>} \left[\left<\right. P^2 \left.\right>
    \left<\right. Q^2 \left.\right> - \left<\right. P,Q
    \left.\right>^2 \right] ,\label{G2}\\
  G^{11}- G^{22} &=& -{i \left<\right. P,Q \left.\right>
    \over \hbar \left<\right. Q^2 \left.\right>},\label{G3}
\end{eqnarray}
where $\left<\right. P,Q \left.\right> \equiv {1\over 2}\left<\right.
(PQ + QP)\left.\right> = (m_0/2)(d/d\tau)\left<\right. Q^2\left.
\right> $ (and $\left<\right. [P,Q]\left.\right> = -i\hbar$).
Substituting $(\ref{q2a}$-$\ref{Qdot^2v})$ gives the values of
$G^{ij}$ at every moment, and the factor $F$ in $(\ref{RDM0})$
will be determined by a normalization condition.

To compare with the transition probability $(\ref{Psho})$, the
reduced density matrix $(\ref{RDM0})$ has to be further transformed to
the representation in the basis of energy-eigenstate for the ``free"
harmonic oscillator $Q$:
\begin{equation}
  \rho^R(Q,Q') = \sum_{m,n\geq 0}\rho_{m,n}^R \phi_m(Q) \phi_n(Q')
\label{rhoeigen}
\end{equation}
where the wave function for the $n$-th excited state is
\begin{equation}
  \phi_n(Q) = \sqrt{\alpha\over 2^n n! \sqrt{\pi}}H_n(\alpha\,Q)
  e^{-\alpha^2 Q^2/2}
\end{equation}
with the Hermite polynomial $H_n(x)$ and the real constant
\begin{equation}
   \alpha = \sqrt{m_0 \Omega_r\over\hbar}.
   \label{alfa}
\end{equation}
Here $\Omega_r$ is the renormalized natural frequency \cite{LH2005}.
Following the method given by Ruiz \cite{Ruiz}, the matrix elements
$\rho^R_{m,n}$ could be extracted by comparing the coefficients of
$s_1^m s_2^n$ on both sides of the following equation,
\begin{equation}
  \sum_{m,n}\sqrt{2^{m+n}\over m! n!}\rho^R_{m,n} s_1^m s_2^n
  = {\alpha \over \sqrt{2 \tilde{G} \left<\right. Q^2 \left.\right>}}
  \exp {1\over\tilde{G}}\left[ \tilde{g}^{11}s_1^2 + \tilde{g}^{22}s_2^2
  +2\tilde{g}^{12}s_1 s_2\right],  \label{rhogen}
\end{equation}
where
\begin{eqnarray}
  \tilde{g}^{11} &=& \left(\tilde{g}^{22}\right)^* = {\alpha^4\over 4} -
   {\left<\right. P^2 \left.\right>\over 4\hbar^2
     \left<\right. Q^2 \left.\right>}
  +{i\alpha^2\left<\right. P,Q \left.\right>\over 2\hbar
     \left<\right. Q^2 \left.\right>}, \\
 \tilde{g}^{12} &=&  = -{\alpha^2\over 2}\left[
   {1\over 4\left<\right. Q^2 \left.\right>}-
    {\left<\right. P^2 \left.\right>\over \hbar^2}
   +{\left<\right. P,Q \left.\right>^2\over \hbar^2
      \left<\right. Q^2 \left.\right>}\right], \\
 \tilde{G} &=& {\alpha^4\over 4} + {\alpha^2\over 2}\left[
   {1\over 4\left<\right. Q^2 \left.\right>}+
    {\left<\right. P^2 \left.\right>\over \hbar^2}
   -{\left<\right. P,Q \left.\right>^2\over \hbar^2
      \left<\right. Q^2 \left.\right>}\right]+
  {\left<\right. P^2 \left.\right>\over 4\hbar^2
     \left<\right. Q^2 \left.\right>} .
\end{eqnarray}

\section{Transition Probability}

The transition probability from the initial ground state to the first
excited state is the $m=n=1$ component of the reduced
density matrix in energy eigenstate representation ($\ref{rhodiag}$):
\begin{equation}
  \rho^R_{1,1} = { \hbar \left[\left<\right. P^2 \left.\right>
    \left<\right. Q^2 \left.\right>
  -\left<\right. P,Q \left.\right>^2-(\hbar^2/4)\right]
  \over \left\{ \left[ \hbar^{-1} \alpha^{-2}\left<\right. P^2 \left.\right>
    + (\hbar /2)\right]
  \left[\hbar \alpha^2\left<\right. Q^2 \left.\right>+ (\hbar/ 2) \right]-
    \left<\right.  P, Q \left.\right>^2\right\}^{3/2}}.
\end{equation}
Expanding the two-point functions of the detector 
$(\ref{q2a}$-$\ref{Qdot^2v})$ in terms of $\gamma$, $\left<\right. Q^2
\left.\right>$ and $\left<\right. P^2 \left.\right>$ look like
\begin{eqnarray}
  \left<\right. Q^2 \left.\right> &\approx& {\hbar\over 2 m_0\Omega_r} +
    \gamma \left<\right. Q^2\left.\right>_{(1)} + O(\gamma^2),\nonumber\\
  \left<\right. P^2 \left.\right> &\approx& {\hbar\over 2}m_0\Omega_r +
    \gamma \left<\right. P^2\left.\right>_{(1)} + O(\gamma^2),
\end{eqnarray}
and $ \left<\right.P,Q \left.\right> \sim O(\gamma )$, which yield
\begin{equation}
  \rho^R_{1,1} \approx {\gamma\over 2 \hbar m_0 \Omega_r}\left[
  \left<\right.P^2\left.\right>_{(1)} + m_0^2 \Omega_r^2
  \left<\right.Q^2\left.\right>_{(1)}\right] + O(\gamma^2) .
\label{pertproba}
\end{equation}
When $\eta\equiv \tau-\tau_0 \gg a^{-1}$ the approximate value up to
the first order of $\gamma\equiv\lambda_0^2/8\pi m_0$ becomes
\begin{equation}
  \rho^R_{1,1}|_{\gamma\eta\to 0} \stackrel{\eta\gg a^{-1}}{\longrightarrow}
  {\lambda_0^2\over 4\pi m_0}\left[ {\eta\over e^{2\pi\Omega_r/a}-1}
  + {\Lambda_1 +\Lambda_0-2\ln (a/\Omega_r) \over 2\pi \Omega_r} \right]
\label{rhopert}
\end{equation}
from $(\ref{q2a}$-$\ref{Qdot^2v})$. Here $\Lambda_0$ and $\Lambda_1$
are large constants in $(\ref{Q^2v})$ and $(\ref{Qdot^2v})$:
$\Lambda_1$ denotes the time resolution/frequency cut-off of this
detector theory, while $\Lambda_0$ denotes  the time scale of
switching on the interaction.

\subsection{Range of validity in conventional results}

We see that the first term of ($\ref{rhopert}$) gives the conventional
transition probability $(\ref{Psho})$ from TDPT over infinite time.
Only when $\Omega_r\eta \gg \Lambda_1, \Lambda_0$, or $a$ is extremely
large, can the second term in $(\ref{rhopert})$ be neglected and
$(\ref{Psho})$ recovered. Hence the conventional transition
probability $(\ref{Psho})$ is valid only in the limits of (a)
ultra-high acceleration ($a\gg \Omega_r$ and $\Lambda_1 \ll a\eta \ll
a\gamma^{-1}$) or (b) ultra-weak coupling  ($a^{-1},
\Omega_r^{-1}\Lambda_1 \ll \eta \ll \gamma^{-1}$).

Note that, in obtaining ($\ref{rhopert}$), we have assumed $a^{-1} \ll
\eta \ll \gamma^{-1}$, when the system is still in transient. If $a <
\gamma$, the conventional transition probability ($\ref{Psho}$) has no
chance to dominate at all. Mathematically, the first term in
$(\ref{rhopert})$ is contributed by the poles at $\pm\Omega+i\gamma$ or
$\pm\Omega-i\gamma$ in the $\kappa$-integrations
of two-point functions (see Eqs.(60) and (67) in \cite{LH2005}).
But when $a < \gamma$ the poles at $\pm i a$ would be closer to the
real $\kappa$ axis than the poles at $\pm \Omega \pm i\gamma$, while
the poles on the imaginary axis $\kappa = \pm i n a$, $n\in N$ become
very dense for small $a$, so their contributions dominate the result
and the first term of $(\ref{rhopert})$ becomes unimportant.

In particular, the $a=0$ case is beyond the reach of TDPT over
infinite time shown in Section I, and the conventional wisdom from
perturbation theory that no transition occurs in an inertial detector
is untenable. In contrast, our result indicates that the evolution of
$\rho^R_{1,1}$ with $a=0$ behaves qualitatively similar to those
cases with nonzero acceleration \cite{LH2005}. This agrees with the
expectation from the observation that the UD detector theory is a
special case of quantum Brownian motion \cite{HM}, where there is
nontrivial interplay between the oscillator and the quantum field at
zero temperature.

\subsection{Essential roles of $\Lambda_0$ and $\Lambda_1$}
\label{Lam1}

We see the presence of two constants $\Lambda_0, \Lambda_1$ and
may wonder whether they have some real physical meaning or are
just part of a calculational tool or artifact. We will address
these concerns here. To begin with, the presence of large
constants corresponding to frequency cut-offs in the coincidence
limit of two-point functions of the detector is a common feature
for detector-field theories and quantum Brownian motion. For
example, the Raine-Sciama-Grove(RSG) model in (1+1)D \cite{RSG},
in which the detector acts like a harmonic oscillator in a
sub-Ohmic bath, also has two-point functions of the detector with
dependence of large constants due to the infrared cut-off.

\begin{figure}
\includegraphics[width=8cm]{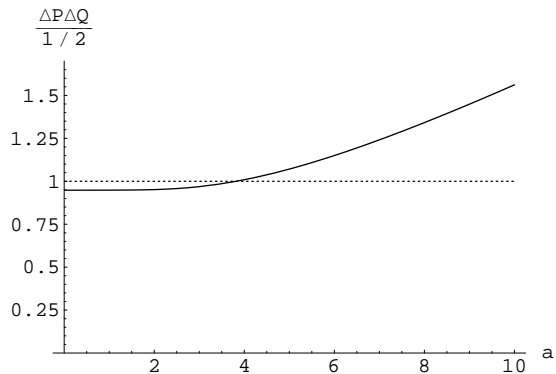}
\caption{Plot of $(\Delta P\Delta Q)/(\hbar/2)$ at late times with
$\Lambda_1=0$. One can see that there exists a range of $a$
(less than about $4$) where $\Delta P\Delta Q \equiv \sqrt{
\left<\right. P^2\left.\right>\left<\right. Q^2\left.\right>}$
is smaller than $\hbar/2$, violating the uncertainty principle.
So $\Lambda_1$ cannot be naively subtracted.
Other parameters in this plot are taken to be $\gamma=0.1$,
$\Omega=2.3$, $m_0=1$, and $\hbar=c=1$.} \label{vioUP}
\end{figure}

Let us explore the physical meaning of $\Lambda_0$ and $\Lambda_1$.
Since $\Lambda_0$ corresponds to the time scale of switching on the
interaction, it could be finite in real processes, and for every
finite value of $\Lambda_0$, the $\Lambda_0$ terms in all two-point
functions vanish at late times (see $(\ref{Q^2v})$ and
$(\ref{Qdot^2v})$). Hence $\Lambda_0$ will not be present in the
late-time results. On the other hand, $\Lambda_1$ is a constant of
time, appearing from the very beginning and never decays. One way to
see that it is a quantity of real physical meaning is that if
$\Lambda_1$ was subtracted naively, the uncertainty principle will be
violated, namely, $\Delta P\Delta Q =\sqrt{\left<\right. P^2
\left.\right> \left<\right. Q^2 \left.\right>}< \hbar/2$ at late
times for $a$ is small enough, as shown in FIG.\ref{vioUP}.

Actually $(\ref{pertproba})$ is formally identical to the
first-order transition probability from TDPT for a UAD with
finite duration of interaction $(\tau_0,\tau)$ \cite{SS92},
so is the result $(\ref{rhopert})$.
To verify this, first note that the $O(\gamma)$ term in
$(\ref{pertproba})$ is the $O(\lambda_0^2)$ term of
\begin{equation}
  {1 \over \hbar \Omega_r}\left< E_0,
  0_M\right.| \hat{H}_Q(\eta)| \left.E_0, 0_M\right>.
\end{equation}
Here $\hat{H}_Q(\eta)$ is defined by
\begin{equation}
  \hat{H}_Q(\eta) \equiv e^{{i\over\hbar}
  \int_{\tau_0}^\tau ds \hat{H}(s)} \hat{H}_Q
  e^{-{i\over\hbar}\int_{\tau_0}^\tau ds \hat{H}(s)},
\end{equation}
where $\hat{H}(\tau)$ is the total Hamiltonian for the combined
system, $\hat{H}_Q$ is the Hamiltonian for the ``free" detector
so that $|\left.E_0\right>$ is an eigenstate of $\hat{H}_Q$.
Then in the interaction picture it is straightforward to show that
\begin{equation}
  \rho^R_{1,1} \approx {\lambda_0^2\over\hbar\Omega_r}
   \int_{\tau_0}^\tau d\tau_1\int_{\tau_0}^\tau d\tau_2 (E_1-E_0)
   \left| \left< E_1\right.| \hat{Q}(0)| \left.E_0\right> \right|^2
   e^{-{i\over\hbar}(E_1-E_0)(\tau_1-\tau_2)}
 \left<\right. 0_M | \Phi(z(\tau_1))\Phi(z(\tau_2))|0_M\left.\right>,
\end{equation}
from which one recovers the finite-time transition probability from
TDPT (cf. Eq.(\ref{Planck})) since $E_1-E_0=\hbar\Omega_r$.
Mathematically the large constants $\Lambda_0$ and $\Lambda_1$ are
formally the same as the divergences found in Ref. \cite{SS92}.
In Ref. \cite{HMP93} it has been shown that these divergences can be
tamed if one switches on and off the interaction smoothly, so can
$\Lambda_0$. Nevertheless, the physical meaning of $\Lambda_1$ here
is totally different from those divergences of the finite-time UAD.
Here we are looking at the real-time causal evolution problem
(``in-in" formulation) rather than a scattering transition amplitude
(``in-out" formulation) problem. Also in our set-up we never turn off
the coupling and $\Lambda_1$ is present at every moment.

As was already found in Ref. \cite{UZ89}
the constant $\Lambda_1$ changes the scenario of the evolution of the
system in an essential way. $\Lambda_1$ is present whenever the
coupling is on, so right after the initial moment the values of
$\rho^R_{m,n}$, $m,n>0$ jump from zero to large numbers depending on
$\Lambda_1$, in a time scale indistinguishable from zero at the level
of precision of this theory. This means that the initial distribution
of $\rho^R_{m,n}$ peaked at the element $\rho^R_{0,0}$ would upon the
switch-on of the coupling collapse rapidly (rather than smoothly
diffuse) into a distribution widely spread (if $\Lambda_1$ is large)
over the whole density matrix. After that the density matrix begins
to redistribute itself and finally settles into a steady state.

\section{Purity, Entropy and Effective Temperature in Detector}

After substituting the late-time two-point functions $(\ref{QQsat})$ and
$(\ref{PPsat})$ with $\Lambda_1$ kept in $\left<\right.P^2\left.\right>$
into $(\ref{rhogen})$, we find that, in steady
state, the diagonal terms of $\rho^R_{m,n}$ in the free eigenstate
representation do not assume the form of a Boltzmann distribution,
while the off-diagonal terms of $\rho^R_{m,n}$ do not vanish at late
times. Therefore the free eigenstates are no longer orthogonal
when interactions are introduced.

To define an effective temperature and the entropy of
entanglement, one should choose another set of orthogonal states
which have a similar form as the energy eigenstate of the free
detector but with $\alpha$ in $(\ref{alfa})$ replaced by
\begin{equation}
  \alpha = \left[{\left<\right. P^2 \left.\right> \over \hbar^2
    \left<\right. Q^2 \left.\right> }\right]^{1/4}. \label{newalpha}
\end{equation}
Since our initial state is Gaussian, one can further apply a canonical
transformation introduced by Unruh and Zurek \cite{UZ89} to diagonize
the reduced density matrix into
\begin{equation}
  \tilde{\rho}^R_{m,n} = \hbar \delta_{mn}{
   \left( {\cal U}- \hbar/2\right)^{n}\,\,\,\,\,
  \over \left({\cal U} + \hbar/2\right)^{n+1}}, \label{diagrho}
\end{equation}
where
\begin{equation}
  {\cal U} \equiv \sqrt{ \left<\right. P^2 \left.\right>
    \left<\right. Q^2 \left.\right> - \left<\right. P,Q
    \left.\right>^2 }.
\end{equation}
is the uncertainty function and ${\cal U}\ge \hbar/2$ is the
Robertson-Schr\"odinger uncertainty relation.
From $(\ref{diagrho})$ one immediately obtains the
purity ${\cal P}$ for the reduced density matrix $(\ref{RDM0})$,
\begin{equation}
  {\cal P} \equiv Tr[ \tilde{\rho}^R \tilde{\rho}^R ] =
  {\hbar/2\over {\cal U}}. \label{purity}
\end{equation}
In this case ${\cal P}^{-1}$ is the ratio of the uncertainty
function ${\cal U}$ to its minimal value $\hbar/2$, hence one
always has ${\cal P} \in (0,1)$ because of the uncertainty
principle. Also the von Neumann entropy or the entropy of
entanglement reads
\begin{equation}
  {\cal S} \equiv - Tr \tilde{\rho}^R \ln \tilde{\rho}^R =
      \left({{\cal U}\over\hbar}+{1\over 2}\right)
      \ln \left({{\cal U}\over\hbar}+{1\over 2}\right) -
      \left({{\cal U}\over\hbar}-{1\over 2}\right)
      \ln \left({{\cal U}\over\hbar}-{1\over 2}\right)
\label{entropy}
\end{equation}
which is a measure of the entanglement between the detector and the
quantum field.

Further, in steady state, $\left<\right. P,Q \left.\right>\to 0$, so
$(\ref{diagrho})$ becomes
\begin{equation}
  \tilde{\rho}^R_{m,n}  \stackrel{\gamma\eta\gg 1}{\longrightarrow}
    \hbar \delta_{mn} {\left[\sqrt{\left<\right. P^2 \left.\right>
    \left<\right. Q^2 \left.\right>}-\hbar/2\right]^{n}\,\,\,\,\,
  \over \left[ \sqrt{\left<\right. P^2 \left.\right>
    \left<\right. Q^2 \left.\right>}+\hbar/2\right]^{n+1}} \equiv
  \delta_{mn}  \rho^R_{0,0}\times e^{-n \hbar\Omega_r /k_B T_{\rm eff}}.
\label{rhodiag}
\end{equation}
Now the diagonal terms assume a Boltzmann distribution with the
effective temperature
\begin{equation}
  T_{\rm eff} = \left[{k_B\over \hbar\Omega_r} \ln
  \left({\sqrt{\left<\right. P^2 \left.\right>
  \left<\right. Q^2 \left.\right>}+\hbar/2  \over
  \sqrt{\left<\right. P^2 \left.\right> \left<\right. Q^2 \left.\right>}-
  \hbar/2 }\right)\right]^{-1}. \label{Teff}
\end{equation}
This is not surprising: Even for such a simple system containing only
two coupled harmonic oscillators, the ground state of the total system
also looks like a thermal state with some effective temperature in view
of the reduced density matrix for one of the oscillator \cite{HKN}.

\subsection{Realistic cases: $\Omega\Lambda_1 \gg a, \gamma$}

\begin{figure}
\includegraphics[width=8cm]{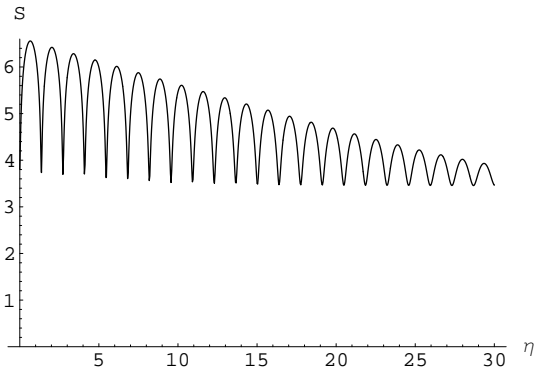}
\caption{The evolution of the entropy of entanglement ${\cal S}(\eta)$
in $(\ref{entropy})$ for a detector initially in the ground state.
At late times, the entropy converges to a large number if
$\Lambda_1$ is large. Here the parameters are taken to be $a=2$,
$\gamma=0.1$, $\Omega=2.3$, $m_0=1$, $\Lambda_1=\Lambda_0=10000$,
and $\hbar=c=1$. } \label{VNS}
\end{figure}

Since $\Lambda_1$ corresponds to the cut-off frequency of the
theory, in real processes it is most likely that $\Omega\Lambda_1 \gg a,
\gamma$. The evolution of the entropy of entanglement in
this case is shown in FIG.\ref{VNS}. At the initial moment the
entropy of entanglement has a sudden jump from zero to a large
number $\sim O( \ln\Lambda_1)$ while the entropy converges to a
number of the same order at late times. Indeed, for $\Lambda_1\gg
1$, the late-time entropy of entanglement has approximately the
value
\begin{equation}
  {\cal S} \approx {1\over 2}\ln \Lambda_1 +1+{1\over 2}\ln \left\{
  {\gamma \over \pi^2 }\left[ {a\over\Omega_r^2}
   + {i\over\Omega} \psi\left( 1+{\gamma-i\Omega\over a}\right)
   - {i\over\Omega} \psi\left( 1+{\gamma+i\Omega\over a}\right)\right]
   \right\} + O(\Lambda_1^{-1}), \label{L1entropy}
\end{equation}
which is dominated by the $\ln \Lambda_1$ term (the $\Lambda_0$
term dies out at late times as long as $\Lambda_0$ is finite).
This indicates that the constant $\Lambda_1$ is due to the
entanglement between the detector and the infinitely many degrees
of freedom of the field.

For $\Lambda_1\gg 1$, the effective temperature is approximately
\begin{equation}
  T_{\rm eff} \approx {\hbar \Omega_r \over \pi k_B}
  \sqrt{\gamma\Lambda_1}\left[ {a \over \Omega_r^2} +
  {i\over\Omega}\psi\left(1+{\gamma-i\Omega\over a}\right)-
  {i\over\Omega}\psi\left(1+{\gamma+i\Omega\over a}\right)\right]^{1/2}
  +O(\Lambda_1^{-1/2}). \label{TeffL}
\end{equation}
which, as we can see,  is totally different from the Unruh
temperature: It is determined not only by the proper acceleration
$a$, but also by the properties of the detector ($\Omega_r$), the
interaction ($\gamma$), and the frequency cutoff of this theory
($\Lambda_1$). 
Moreover $T_{\rm eff}$ is very large ($O(\sqrt{\Lambda_1}))$ for
large $\Lambda_1$ and non-vanishing even when $a\to 0$.

\subsection{Inertial detector: $a=0$}

Substituting $(\ref{Q^2a0r0})$ and $(\ref{P^2a0r0})$ into
$(\ref{purity})$, $(\ref{entropy})$ and $(\ref{Teff})$, one can
write down in closed form the purity, entropy of entanglement and
the effective temperature of an inertial detector ($a=0$). These
quantities are still nontrivial. In particular, in the ultra-weak
coupling limit $\gamma \Lambda_1 \ll 1$, one has
\begin{eqnarray}
  {\cal S} &\approx& {\gamma(\Lambda_1-1)\over \pi\Omega_r}\left(1-
    \ln{\gamma(\Lambda_1-1)\over\pi\Omega_r}\right) + O(\gamma^2),\\
  T_{\rm eff} &\approx& {\hbar\Omega_r\over - k_B
  \ln\left[{\gamma(\Lambda_1-1)\over \pi\Omega_r}+ O(\gamma^2)\right]}
\end{eqnarray}
[Note that $\lim_{\epsilon \to 0^+}\ln (-1+\epsilon) = -i\pi$).]
Both quantities go to zero as $\gamma \to 0$, when the description
of the combined system returns to the nearly free theory.

\subsection{Conventional result: the Markovian regime}

As we mentioned in Section III.A, the transition probability
calculated from TDPT over infinite time
is valid only in the ultra-high acceleration
(temperature) limit or the ultra-weak coupling limit, both
corresponding to the Markovian regime for this model (with the
quantum field as an Ohmic bath). In these limits the entropy of
entanglement and effective temperature do have the approximate
values expected in the conventional picture:

(a) {\it Ultra-high acceleration (temperature) limit,}
$a\gg \gamma\Lambda_1,\Omega$. In this limit, the late-time entropy
of entanglement reads
\begin{equation}
  {\cal S} \approx \ln {a\over 2\pi\Omega}+1 + O(a^{-1}),
\end{equation}
and the effective temperature
\begin{equation}
  T_{\rm eff} \approx T_U + {\hbar\gamma\over k_B\pi}\left( \Lambda_1 +
  \gamma_e -\ln{a\over\Omega}\right) + O(a^{-1}),
\end{equation}
is very close to the Unruh temperature $T_U= \hbar a/2\pi k_B$.

(b) {\it Ultra-weak coupling limit,} $\gamma \Lambda_1 \ll a,\Omega$.
In this limit, the late-time entropy of entanglement and effective
temperature read
\begin{eqnarray}
  {\cal S} &\approx& {\pi\Omega\over a}\coth {\pi\Omega\over a} -
    \ln \left[ 2\sinh{\pi\Omega\over a}\right] +{2\gamma\over a}\left\{
    \Lambda_1 -\ln{a\over\Omega} - {\rm Re}\,\left[
    \psi\left({i\Omega\over a}\right)+ {i\Omega\over a}\psi^{(1)}
    \left({i\Omega\over a}\right)\right]\right\} + O(\gamma^2),\\
  T_{\rm eff} &\approx& T_U + {\hbar \gamma a^2\over k_B \pi^3 \Omega^2}
    \sinh^2{\pi\Omega\over a}\left\{ \Lambda_1 -\ln{a\over\Omega}
    - {\rm Re}\,\left[ \psi\left({i\Omega\over a}\right)+ {i\Omega\over a}
    \psi^{(1)}\left({i\Omega\over a}\right)\right]\right\}+O(\gamma^2).
\end{eqnarray}
Again the effective temperature of the detector is very close to
the Unruh temperature.

It is well-known that the quantum mechanical time-dependent
perturbation theory invokes the equivalent of a Markovian
approximation (e.g., as in the derivation of the Pauli master
equation \cite{Reif}). Indeed, both the above limits correspond to
the Markovian regime of the quantum Brownian motion, its dynamics
being described by the master equation of Caldeira and Leggett
\cite{CalLeg83} (in contradistinction to the non-Markovian master 
equation of Hu, Paz, and Zhang \cite{HPZ}). There is admittedly a
fluctuation-dissipation relation between the fluctuations in the
field and the dynamics of the detector and thus backreaction is
present, but with ultra-weak coupling it is restricted to the
linear response regime (which has limited domain of applicability
compared to the full nonequilibrium dynamics of the combined
system). Here, the thermal bath is only slightly affected by the
back reaction from the detector to the field, namely, the detector
acts essentially as a test particle in the field.
It is only under these special assumptions that the detector can
come to equilibrium with an approximate thermal field at the
Unruh temperature. Beyond this regime, the dynamics of the
detector-field system has a totally different character.

\section{Remarks}

We conclude with a few remarks:\\

\paragraph{Working range of Unruh-DeWitt detectors.}
Conventional results from time-dependent perturbation theory over an
infinitely long duration of interaction are trustworthy only in the
ultra-high acceleration (or temperature) limit and the ultra-weak
coupling limit, under the Markovian regime. In the non-Markovian
regimes, the large constants $\Lambda_1$ and $\Lambda_0$ alter the
scenario fundamentally. With the presence of the large constant
$\Lambda_1$, which reflects the entanglement between the detector and
the infinitely many degrees of freedom of the field, the evolution of
the reduced density matrix of the detector collapses (sudden rapid
change), as shown here, rather than diffuses (smooth gradual change),
as conjured in the conventional picture. Furthermore, the detector at
late times never sees an exact Boltzmann distribution over the energy
eigenstates of the free detector, let alone the Unruh temperature.
The strong interplay between the detector and the field makes the
late-time quantum states for the detector-field system highly
entangled and never factorizable (an analogy in quantum optics is the
photon-atom bound state \cite{JQ94}.)

In equilibrium conditions, one can diagonalize the reduced density
matrix and define an effective temperature $(\ref{TeffL})$, which is
usually large ($\sim O(\sqrt{\Lambda_1})$) and does not vanish even
for an inertial detector in Minkowski vacuum as long as the coupling
is on. Only in the Markovian regime indicated above could this
effective temperature get very close to the Unruh temperature.
\\

\paragraph{Correspondence with QBM.}

For a uniformly accelerated UD detector in (3+1)D with proper
acceleration $a$ the Unruh effect \cite{Unr76, DeW79, BD} attests
that it should behave the same way as an inertial UD detector in
contact with a thermal bath at Unruh temperature $T_U$, or more
precisely, as an inertial harmonic oscillator in contact with an
Ohmic bath at $T_U$ \cite{FHR}. This is clear from an examination of
the integral in the derivation of the two-point functions of the
detector, for example, Eq.(60) in Ref.\cite{LH2005}:
\begin{equation}
  \left<\right. Q(\eta)Q(\eta')\left.\right>_{\rm v} \sim
  {\lambda_0^2 \hbar\over (2\pi)^2 m_0^2 } \int
  {\kappa d\kappa\over 1-e^{-2\pi \kappa/a}}[\ldots], \label{planck}
\end{equation}
Therefore for an inertial UD detector in contact with an Ohmic bath
at temperature $T$, one can simply substitute $2\pi k_B T/\hbar$ for
$a$ in the above results to get its  purity, entropy and effective
temperature.

However, examining this from the vantage point of the exact solutions
we obtained,  we see the above statements are accurate only at the
initial moment. After the coupling is switched on, the quantum state
of the field will have been changed by the detector, so the field is
no longer in the Minkowski vacuum and it does not make exact sense to
say that the detector is immersed in a thermal state (or any state
defined in the test-field description, i.e., where the field is
assumed not to be modified by the presence of the detector).

A theorem by Bisognano and Wichmann (BW) \cite{BW75} states that the
Minkowski vacuum, which is uniquely characterized by its invariance 
under all Poincar\'{e} translations, is a Kubo-Martin-Schwinger (KMS) 
state with respect to all observables confined to a Rindler wedge. 
One may wonder why it does not apply here. The reason is that the BW 
theorem refers to the vacuum state of a quantum field alone, not the 
combined detector-field system. Even when the combined system is in 
a steady state, the quantum state of the interacting field is not       
invariant under spatial translations in Minkowski space, hence 
is not covered under the assumption of the BW theorem pertaining to 
Poincar\'{e} invariance.

Actually the Planck factor in $(\ref{planck})$ is a consequence of
the BW theorem. Nevertheless, it is derived from only the
free-field-solution part of the complete interacting field  (see Eqs.
(28), (56) and (58) in Ref. \cite{LH2005}). Here the factor is not
distorted by the interaction simply because the field is linear and
the coupling is bilinear. For nonlinear fields or couplings it would
have a nonPlanckian spectrum and the departure from the conventional 
picture would be more pronounced.                                    
\\

\paragraph{Backreaction and memory effects.}  It is common
knowledge in nonequilibrium statistical mechanics \cite{Zwanzig} that
for two interacting subsystems the two ordinary differential
equations governing each subsystem can be written as an
integro-differential equation governing one such subsystem, thus
rendering its dynamics non-Markovian, with  the memory of the other
subsystem's dynamics registered in the nonlocal kernels (which are
responsible for the appearance of dissipation and noise should the
other subsystem possesses a much greater number of degrees of freedom
and are coarse-grained in some way). Thus inclusion of back-action
self-consistently in general engenders non-Markovian dynamics. For
our problem the two subsystems are the detector and the quantum
field.  Combining Eqs. (28), (30) and (10) in Ref.\cite{LH2005}, we
can write down the equation of motion for the detector evolution
functions,
\begin{equation}
  (\partial_\tau^2 + \Omega_0^2)q^{(+)}(\tau;{\bf k})=
  {\lambda_0\over m_0} \left[ f^{(+)}_0(z(\tau);{\bf k})+ {\lambda_0}
  \int_{\tau_0}^\infty d\tau' G_{\rm ret}(z(\tau);z(\tau'))
  q^{(+)}(\tau';{\bf k})\right],
\end{equation}
which is an integro-differential equation. The backreaction to the
detector is registered through the retarded Green's function $G_{\rm
ret}$ of the field. Various approximations are usually invoked to
solve this equation, amongst which the most common is the Markovian
or memoryless approximation. This is what enters in the conventional
derivation of the Unruh effect, but as we see above, 
is a very special and nongeneric subcase.\\

\paragraph{Hiding $\Lambda_1$?}
One may wonder why the large constant $\Lambda_1$ cannot be absorbed
by any parameter of this theory so one can renormalize $T_{\rm
eff}$. To begin with, as we mentioned in Sec.\ref{Lam1}, the UD
detector theory is not a fundamental theory to meet the
renormalizability requirement, and the presence of cut-offs as
physical parameters is an expected feature which characterizes the
range of validity of this semiclassical theory, just like the
 Compton wavelength of the electron acting as a cut-off in quantum
optics. Second, the large constant $\Lambda_1$ is not present in the
renormalized stress-energy tensor of the field induced by the
detector \cite{LH2005}, thus $T_{\rm eff}$ may not be a directly
measurable quantity. 
The interference between the vacuum fluctuations and their back
reaction cancels this cut-off dependence so that $\Lambda_1$ is not
observable outside of the detector.

In some cases, though, $\Lambda_1$ term can be subtracted from the
physical quantities of the detector {\it after} these quantities are 
worked out. For example, one can subtract $\Lambda_1$ from the 
detector energy defined in Eq.(82) of Ref.\cite{LH2005} because 
during a physical process only the difference of energy matters.\\

\paragraph{Hawking effect.}
It is tempting to see if analogous descriptions can be made and
implications drawn for the Hawking effect. For a UD
detector at rest in a static gravitational field, the response of
the detector is similar. The simplest example is the UD detector
fixed at radius $r$ far from a Schwarzschild black hole, while the
scalar field is in Hartle-Hawking (HH) vacuum (which is the
counterpart of Minkowski vacuum for the uniformly accelerated
detector in Minkowski space \cite{BD}). The response function per
unit time for a massless scalar field in the Schwarzschild
background is given by Candelas (Section V in Ref.\cite{Cande80}).
According to \cite{Cande80}, when $r\to \infty$, the response
function for HH vacuum is exactly the same as the one
in the uniform acceleration case (Eq.(58) in \cite{LH2005}) with
the Unruh temperature $a/2\pi k_B$ replaced by the Hawking
temperature $T_H \equiv 1/8\pi k_B M$ for the Schwarzschild black
hole with mass $M$. Therefore the results in this paper can be
directly applied to the case of UD detector fixed at $r\to \infty$
outside a Schwarzschild black hole by neglecting its back reaction
to spacetime.
The effective temperature $T_{\rm eff}$ read off from such a
detector in steady state is given by $(\ref{Teff})$ with $a$
replaced by $1/4M$ rather than the Hawking temperature $T_H$.

For the case with the detector sitting very close to the event
horizon, due to the effective potential barrier in the radial
equation for the field, one has to consider the back-scattering of
the retarded field induced by the detector in addition to the vacuum
fluctuations described by the response function. We expect that the
effective temperature read off from the detector in steady state
would be even more complicated but definitely different from $T_H$.
We hope to report on this investigation in a
later paper.\\

\noindent{\bf Acknowledgement} We wish to thank Chris Fleming and
Albert Roura for discussions, Bill Unruh for pointing out his paper
with Zurek, and Ted Jacobson and Ralf Schutzhold helpful suggestions
in presenting our results. This work is supported in part by NSF
grant PHY-0601550.

\begin{appendix}
\section{Two-Point Functions of Inertial Detectors in Minkowski Vacuum}

Recall that for a uniformly accelerated UD detector with proper
acceleration $a$ \cite{LH2005}, once we choose the factorized
initial state $(\ref{initstat})$, the two-point functions  split
into two part, $\left<\right.\ldots\left.\right>=
\left<\right.\ldots\left.\right>_{\rm a}+
\left<\right.\ldots\left.\right>_{\rm v}$. The coincidence limits
of the two-point functions of the detector with respect to its
ground state (with $q_0$ in Ref.\cite{LH2005} being zero) read
\begin{eqnarray}
\left<\right.Q(\eta)^2\left.\right>_{\rm a} &=&
  {\hbar \theta(\eta)\over 2\Omega^2\Omega_r m_0}e^{-2\gamma\eta}
  \left[\Omega_r^2 - \gamma^2\cos 2\Omega\eta +
  \gamma\Omega\sin 2\Omega\eta\right],\label{q2a}\\
\left<\right.\dot{Q}(\eta)^2\left.\right>_{\rm a} &=&
  {\hbar \Omega_r \over 2\Omega^2 m_0}\theta(\eta)e^{-2\gamma\eta}\left[
  \Omega_r^2 -\gamma^2\cos 2\Omega\eta -\gamma\Omega\sin 2\Omega\eta\right],
  \label{dotQ2a}
\end{eqnarray}
which are independent of $a$, while the coincidence limits of the
two-point functions of the detector with respect to Minkowski
vacuum read
\begin{eqnarray}
\left<\right. Q(\eta)^2\left.\right>_{\rm v}
  &=&  \lim_{\eta'\to\eta} {1\over 2}\left<\right. \left\{ Q(\eta),
    Q(\eta')\right\}\left.\right>_{\rm v} \nonumber\\
  &=& {2\hbar\gamma  \over \pi m_0 \Omega^2} \theta(\eta){\rm Re}\left\{
  \left(\Lambda_0 - \ln {a\over\Omega}\right)
    e^{-2\gamma\eta}\sin^2\Omega\eta \right. \nonumber\\ &+& {a\over 2}
    e^{-(\gamma +a)\eta}\left[{F_{\gamma+i\Omega}(e^{-a\eta})\over \gamma+i
    \Omega+a}\left( -{i\Omega\over\gamma}\right)e^{-i\Omega\eta}
   + {F_{-\gamma - i\Omega}(e^{-a\eta})\over \gamma + i\Omega-a}\left(
    \left(1+{i\Omega\over\gamma}\right)e^{i\Omega\eta} -e^{-i\Omega\eta}\right)
    \right]\nonumber\\
  &-& {1\over 4}\left[ \left({i\Omega\over\gamma}+ e^{-2\gamma\eta}
  \left({i\Omega\over\gamma}+1 -e^{-2i\Omega\eta}\right)\right)
  \left( \psi_{\gamma+i\Omega}+ \psi_{-\gamma-i\Omega}\right)\right.\nonumber\\
  & & \left.\left. - \left(-{i\Omega\over\gamma}+e^{-2\gamma\eta}
  \left({i\Omega\over\gamma}+1 -e^{-2i\Omega\eta}\right)\right)i\pi\coth
  {\pi\over a}(\Omega-i\gamma)\right]\right\},\label{Q^2v}\\
\left<\right. \dot{Q}(\eta)^2\left.\right>_{\rm v}
    &=& {2\hbar\gamma  \over \pi m_0 \Omega^2}
  \theta(\eta){\rm Re}\,\left\{\left(\Lambda_1-\ln {a\over\Omega}\right)\Omega^2 +
  \left(\Lambda_0-\ln {a\over\Omega}\right)e^{-2\gamma\eta}\left(\Omega
   \cos\Omega\eta-\gamma \sin\Omega\eta \right)^2
   \right. \nonumber\\
  &+& {a\over 2}(\gamma+i\Omega)^2 e^{-(\gamma +a)\eta} \left[{F_{\gamma+i\Omega}
    (e^{-a\eta}) \over \gamma+i\Omega+a}\left( {i\Omega\over\gamma}\right)
    e^{-i\Omega\eta} + {F_{-\gamma-i\Omega}(e^{-a\eta})\over \gamma +i\Omega-a}\left(
    \left(1-{i\Omega\over\gamma}\right)e^{i\Omega\eta} -e^{-i\Omega\eta}\right)
    \right]\nonumber\\
  &+& {1\over 4} (\gamma +i\Omega)^2\left[\left( {i\Omega\over\gamma}+
    e^{-2\gamma\eta}\left({i\Omega\over\gamma}-1+e^{-2i\Omega\eta}\right)\right)
  \left(\psi_{\gamma+i\Omega}+  \psi_{-\gamma-i\Omega}\right)\right.\nonumber\\
  & & \left.\left. -\left(-{i\Omega\over\gamma}+e^{-2\gamma\eta}
  \left({i\Omega\over\gamma}-1+e^{-2i\Omega\eta}\right)\right)i\pi\coth
  {\pi\over a}(\Omega-i\gamma)\right]\right\}.\label{Qdot^2v}
\end{eqnarray}
Here again $\eta\equiv \tau-\tau_0$ is the duration of interaction,
$\gamma\equiv\lambda_0^2/8\pi m_0$,
$\Omega \equiv \sqrt{\Omega_r^2-\gamma^2}$,
$F_s(y)$ is defined by the hyper-geometric function as
\begin{equation}
  F_s(y) \equiv {}_2 F_1\left(1+{s\over a},1,2+{s\over a}; y\right),
\end{equation}
and
\begin{equation}
  \psi_s \equiv \psi\left(1+{s\over a}\right)
\end{equation}
is the digamma function. The large constant $\Lambda_0\equiv -\ln
\Omega|\tau_0-\tau_0'|-\gamma_e$ with the Euler's constant $\gamma_e$
corresponds to the time scale of switching-on the interaction, so
$\Lambda_0$ could be finite in real processes, and for every finite
value of $\Lambda_0$, the terms containing $\Lambda_0$ in
$(\ref{Q^2v})$ and $(\ref{Qdot^2v})$ vanish as $\gamma\eta\to\infty$.
The other large constant $\Lambda_1 \equiv -\ln \Omega|\tau-
\tau'|-\gamma_e$ corresponds to the time-resolution or the cut-off
frequency of this theory. Note that here we use  slightly
different definitions of $\Lambda_1$ and $\Lambda_0$ from those of
Ref.\cite{LH2005} to make the arguments in logarithm functions
dimensionless. Note also that the above results are valid when
$\Omega|\tau-\tau'|\ll 1$ and $\Omega |\tau_0-\tau_0'|\ll 1$.
Beyond this regime, the form of the above ``coincidence" limit of
two-points functions should be modified.

At late times ($\gamma\eta \gg 1$), the two-point functions of the
detector with respect to the initial ground state ($\left<
\ldots\right>_{\rm a}$) die away, and the two-point functions of
the detector saturate to
\begin{eqnarray}
\left<\right. Q^2\left.\right> &\to& {\hbar\over 2\pi m_0\Omega} {\rm Re}\,
  \left[{ia\over \gamma+i\Omega}-2i\psi_{\gamma+i\Omega}\right],
  \label{QQsat}\\
\left<\right. P^2\left.\right> = m_0^2\left<\right. \dot{Q}^2\left.\right>
  &\to& {\hbar m_0\over 2\pi \Omega}
  {\rm Re}\,\left\{ (\Omega-i\gamma)^2 \left[ {ia\over \gamma+i\Omega}-
  2i\psi_{\gamma+i\Omega}\right]\right\} + {2\over \pi}\hbar m_0 \gamma
  \left(\Lambda_1  -\ln {a\over\Omega}\right), \label{PPsat}
\end{eqnarray}
which are identical to the results of the quantum Brownian motion
of a harmonic oscillator in contact with an Ohmic bath at the Unruh
temperature initially \cite{FHR}. To satisfy the uncertainty principle
for all $a$ (see FIG.\ref{vioUP}), here we keep the constant
$\Lambda_1$, which was subtracted in Ref.\cite{LH2005} because of
the observation that $\Lambda_1$ will not be seen outside of the
detector.

When $a\to 0$, the regular terms in the two-point functions
diverge as $\ln a$ and cancel the ones following $\Lambda_0$ and
$\Lambda_1$, so the two-point
functions remain well-behaved. 
In this limit the two-point function $(\ref{Q^2v})$ continuously
approaches
\begin{eqnarray}
  \left<\right. Q(\eta)^2\left.\right>_{\rm v}|_{a=0}
  &=& {\hbar\gamma\over \pi m_0 \Omega^2} \theta(\eta){\rm Re}\left\{
  2\Lambda_0 e^{-2\gamma\eta}\sin^2\Omega\eta \right.
  \nonumber\\ &-& {i\Omega\over\gamma}
  \Gamma\left[0, (\gamma+i\Omega)\eta\right]- e^{-2\gamma\eta} \left(
    {i\Omega\over\gamma}+1-e^{-2i\Omega\eta}\right)
    \Gamma\left[0, -(\gamma+i\Omega)\eta\right]\nonumber\\
  &-& \left. \left({i\Omega\over\gamma}+ e^{-2\gamma\eta}
  \left({i\Omega\over\gamma}+1 -e^{-2i\Omega\eta}\right)\right)
  \ln \left({\gamma\over\Omega}+i\right)
  -\pi e^{-2\gamma\eta}\left({\Omega\over\gamma}+
  \sin 2\Omega\eta\right)\right\},\label{Q^2va0}
\end{eqnarray}
where $\Gamma$ is the incomplete gamma function, while $(\ref{Qdot^2v})$
goes to
\begin{eqnarray}
  \left<\right. \dot{Q}(\eta)^2\left.\right>_{\rm v}|_{a=0}
  &=& {\hbar\gamma\over \pi m_0 \Omega^2} \theta(\eta){\rm Re}\left\{
  2\left[ \Lambda_1 \Omega^2+
   \Lambda_0 e^{-2\gamma\eta}\left(\Omega
   \cos\Omega\eta-\gamma \sin\Omega\eta \right)^2 \right] \right.
  \nonumber\\ &+& \left[{i\Omega\over\gamma}
  \Gamma\left[0, (\gamma+i\Omega)\eta\right]+ e^{-2\gamma\eta} \left(
    {i\Omega\over\gamma}-1+e^{-2i\Omega\eta}\right)
    \Gamma\left[0, -(\gamma+i\Omega)\eta\right]\right.\nonumber\\
  & & \left. + \left({i\Omega\over\gamma}+ e^{-2\gamma\eta}
  \left({i\Omega\over\gamma}-1 +e^{-2i\Omega\eta}\right)\right)
  \ln \left({\gamma\over\Omega}+i\right)\right](\gamma+i\Omega)^2
  \nonumber\\ &+& \left.\pi e^{-2\gamma\eta}\left[-{\Omega\over\gamma}
  (\Omega^2+\gamma^2)+2 \gamma\Omega\cos 2\Omega\eta +
  (\Omega^2-\gamma^2)\sin 2\Omega\eta\right]\right\}.\label{dotQ^2va0}
\end{eqnarray}
At late times, one has
\begin{eqnarray}
  \left<\right. Q^2\left.\right>|_{a=0}&\to&{\hbar i\over 2\pi m_0\Omega}
    \ln {\gamma-i \Omega\over \gamma+ i\Omega}, \label{Q^2a0r0}\\
  \left<\right. P^2\left.\right>|_{a=0} &\to&{\hbar m_0\over\pi}
  \left\{ {i\over 2\Omega}(\Omega^2-\gamma^2)\ln {\gamma-i\Omega\over
    \gamma+i\Omega} +\gamma\left[ 2\Lambda_1 -\ln \left(1+
    {\gamma^2\over\Omega^2}\right)\right]\right\}. \label{P^2a0r0}
\end{eqnarray}
Substituting this into $(\ref{purity})$, $(\ref{entropy})$ and
$(\ref{Teff})$, one obtains the purity, entropy of entanglement
and effective temperature for the inertial detector in the
Minkowski vacuum.

\end{appendix}

\end{document}